\documentclass[12pt, draftclsnofoot, onecolumn]{IEEEtran}
\usepackage{cite}
\usepackage{amsmath,amssymb,amsfonts}
\usepackage{algorithmic}
\usepackage{graphicx}
\usepackage{textcomp}
\usepackage{xcolor}
\usepackage{xfrac}
\usepackage{soul}
\usepackage[caption=false]{subfig}
\begin{document}
\title{Inertial Sensor Aided mmWave Beam Tracking to Support Cooperative
Autonomous Driving}
\author{Mattia Brambilla, Monica Nicoli, Sergio Savaresi, Umberto Spagnolini
\\
 \normalsize
{Politecnico di Milano, Milan, Italy} \\
 E-mails:\{mattia.brambilla, monica.nicoli, sergio.savaresi, umberto.spagnolini\}$@$polimi.it }

\maketitle

\begin{abstract}
	\normalsize
This paper presents an inertial sensor aided technique for beam alignment
and tracking in massive multiple-input multiple-output (MIMO) vehicle-to-vehicle
(V2V) communications based on millimeter waves (mmWave). Since directional
communications in vehicular scenarios are severely hindered by beam
pointing issues, a beam alignment procedure has to be periodically
carried out to guarantee the communication reliability. When dealing
with massive MIMO links, the beam sweeping approach is known to be
time consuming and often unfeasible due to latency constraints. To
speed up the process, we propose a method that exploits a-priori information
on array dynamics provided by an inertial sensor on transceivers to assist the beam alignment procedure. The proposed inertial sensor
aided technique allows a continuous tracking of the beam while transmitting,
avoiding frequent realignment phases. Numerical results based on real
measurements of on-transceiver accelerometers demonstrate a significant
gain in terms of V2V communication throughput with respect to conventional
beam alignment protocols.
\end{abstract}

\begin{IEEEkeywords}
	\normalsize
Beam alignment, beam tracking, inertial sensor, V2V, mMIMO 
\end{IEEEkeywords}

\section{Introduction}

\label{sec:intro}

Connected autonomous vehicles are expected to improve safety, efficiency
and comfort of mobility, disrupting the paradigm of traditional human-controlled
driving \cite{SAE}. Vehicle-to-anything (V2X) communications enable
fast exchange of massive sensor data and mobility patterns between
autonomous vehicles, opening the door to the so-called cooperative
sensing and maneuvering functionalities \cite{Kovacs2015,Brambilla2018,Soatti2018}, which
have been proved to augment the perception capability and the traffic
efficiency. Considered the challenging requirements in terms of latency
and data-rate \cite{3GPP_Rel16_2,Heath2016}, today the millimeter-wave
(mmWave) technology is deemed as the only viable radio frequency approach to support the V2X connectivity, thanks to the large availability of bandwidth
in this spectrum region. Ultra-reliable fifth generation (5G) cellular
standards are currently under development to meet the automotive use case requirements.
They aim at ensuring high data-rate (1 Gbps), ultra-low packet loss
($10^{-7}$) and ultra-low latency (1 ms) for tactile-like safety-critical
applications \cite{Simsek2016}. However, the severe path loss faced
at mmWave frequencies, along with the effects of the atmospheric absorption
and human/environmental obstructions, might significantly hinder the
communication performance if not properly addressed. Mobility, Doppler
effect, blockage and lack of context information are also critical
issues to be considered in vehicular environments \cite{Zorzi2017}.

A solution is to adopt large antenna arrays at both side of the communication
system, so as to form sharp radiation beams and compensate the high
path-loss. Such massive multiple-input multiple-output (mMIMO) systems
are expected to become a pervasive technology in smart mobility applications,
thanks to the feasible array dimension (proportional to the mmWave
wavelength) and moderate energy consumption. This approach, however,
requires precise beam alignment (BA) and tracking procedures to guarantee
the continuity of the communication performance as the devices move.
An exhaustive search of the optimal transmit/receive beam pair might
be too time demanding in vehicular scenarios, considering the latency
constraints. To speed up the BA procedure, different solutions have
been proposed in the literature \cite{Wang2009,Du2013,Heath2014,Nitsche2015}.
For vehicular applications, the authors of \cite{Perfecto2017} propose
to explore the channel and queue state information to optimize both
transmission and reception beamwidths. Other promising approaches
exploit side information to support the communication, such as information
provided by a radar signal operating in a different mmWave band in
\cite{Prelcic2016} or vehicle motion prediction based on global positioning
system (GPS) data in \cite{Mavromatis2017}.

Integration of data by transceiver's sensors is a leading concept
that is expected to be a turning point for the development of vehicle-to-vehicle
(V2V) communications. This paper proposes to use acelerometers on
the array of antennas to optimize the performance of mmWave V2V communications.
In particular, the BA phase carried out by two vehicles before the
data transmission is substituted by a faster signaling of predictive
information on mutual V2V array dynamics. This approach results in
an overhead reduction and it allows a continuous beam steering while
data transmission occurs. The proposed approach applies also to vehicle-to-infrastructure
(V2I) communications, with straightforward adaptations.

\section{Beam Alignment Problem}

\begin{figure}[]
	\centering
	\includegraphics[width=0.7\columnwidth]{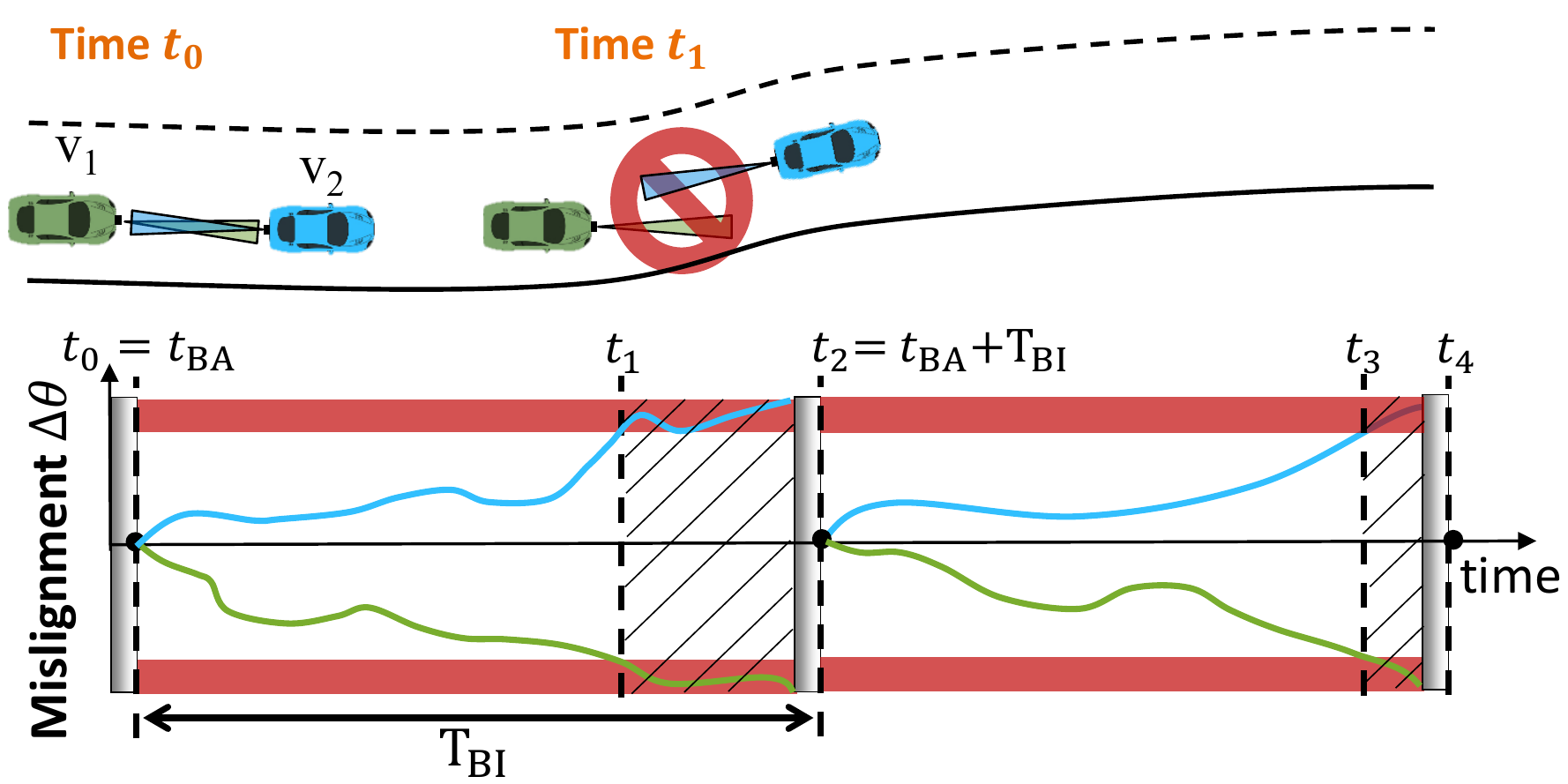}
	\caption{BA problem: at time $t=t_0$ vehicles are correctly pointing their beams. At time $t=t_1$, the communication drops since beams are misaligned due to vehicle dynamics. To avoid this problem, vehicles must periodically perform a BA procedure with period $\mathrm{T}_{\text{BI}}$.}
	\label{fig:BA problem}
\end{figure}

We consider two vehicles $\upsilon_{1}$ and $\upsilon_{2}$ equipped
with mMIMO communication devices on front and rear bumper, respectively,
as illustrated in Fig. \ref{fig:BA problem}. The transceiver on each
vehicle is assumed to sense the information on the array of antennas
dynamic. Let $t=0,1,2,...$ denote the discrete time with sampling
interval $\Delta t$, a perfect alignment is observed right after
the BA procedure at time $t=t_{\text{BA}}$, with the two vehicles
$\upsilon_{1}$ and $\upsilon_{2}$ pointing their transmit/receive
beams towards the line of sight (LOS) direction. As the vehicles move,
V2V connectivity is affected by beam fluctuations due to the relative
vibrations and tilting that can easily lead to communication drops
when sharp beams are employed, as for $t\in[t_{1},t_{2}]$ and $t\in[t_{3},t_{4}]$
in Fig. \ref{fig:BA problem}. To avoid this problem, BA has to be
periodically employed, by a beam sweeping procedure (e.g., exhaustive
or hierarchical \cite{Wang2009}) performed at the beginning of each
beacon interval (BI) \cite{Perfecto2017}, i.e. with repetition time
$\mathrm{T_{BI}}$. The BI length $\mathrm{T_{BI}}$ should be selected
so as to optimize the throughput, as a short $\mathrm{T_{BI}}$ reduces
the effective time for data transfer, while a long one might be inefficient
to track the vehicle dynamics.

Exemplifying the vehicle as a rigid body rotating around its
three axis (longitudinal, perpendicular and lateral), in this work we consider
only the vehicle rotation around the lateral one, i.e. the
chassis pitch, which simplifies the analysis of BA problem relative to the vertical vibrations only. However, the proposed methodology is general enough to be extended to the complete 3D problem, where all the three rotations contribute to determine the vehicle attitude. Here, we 
analyze the impact of vertical vehicle oscillations (i.e., strokes)
on the V2V communication between vehicles $\upsilon_{1}$ and $\upsilon_{2}$
and we propose a method to improve the beam pointing based on stroke
measurements provided by an inertial sensor on antennas, as discussed
in the following. 
\begin{figure}[b]
	\centering
	\includegraphics[width=0.7\columnwidth]{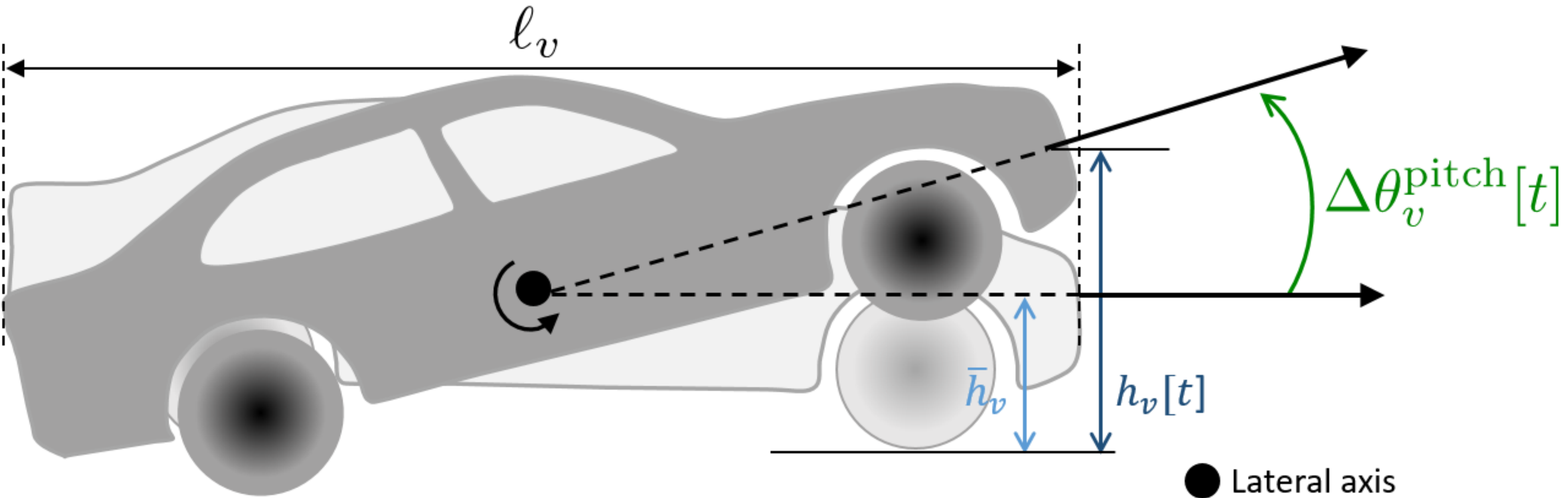}
	\caption{Vehicle pitch from measurement of front height variations by a sensor inside the communication device.}
	\label{fig:rotation}
\end{figure}
As illustrated in Fig. \ref{fig:rotation}, let $\bar{h}_{v}$ be the height of the antenna array in rest conditions
for vehicle $v$ and $h_{v}[t]$ the height observed by an inertial
sensor at time $t$, the pitch angle of vehicle $v$ at time $t$
is approximated by: 
\begin{align}
\Delta\theta_{v}^{\mathrm{pitch}}[t]=\text{tan}^{-1}\bigg(\frac{h_{v}[t]-\bar{h}_{v}}{0.5\ell_{v}}\bigg),\label{eq:pitch angle}
\end{align}
where $\ell_{v}$ indicates the vehicle length. The pitch variations
over time impact on the V2V communication link, as shown in Fig. \ref{fig:Geometry}.
\begin{figure}
	\centering
	\includegraphics[width=0.7\columnwidth]{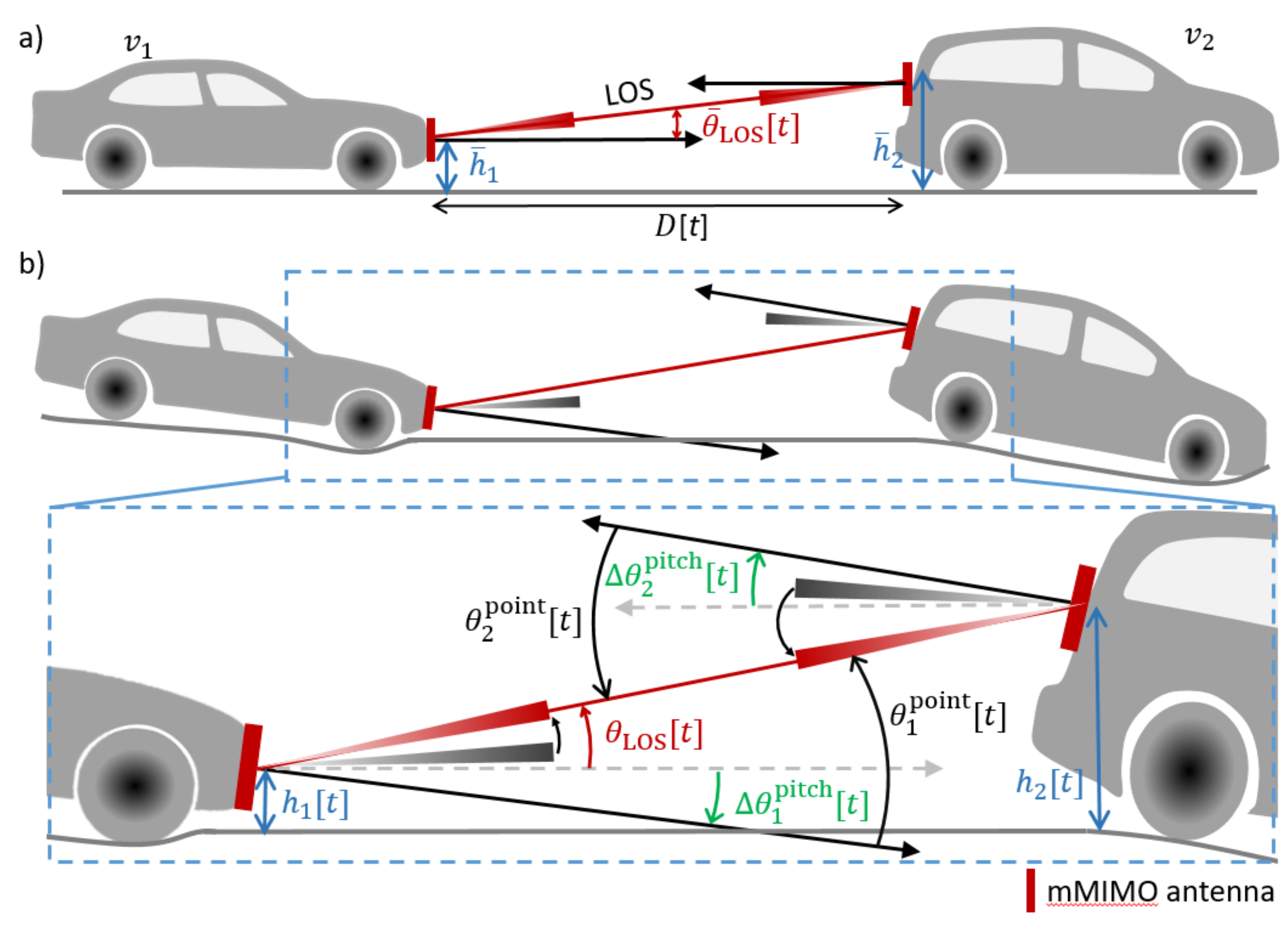}
	\caption{a) V2V communication geometry in a rest condition. b) Impact of the chassis pitch with emphasis on beam steering.}
	\label{fig:Geometry}
\end{figure}

In ideal conditions with null pitch, i.e. for $\Delta\theta_{1}^{\mathrm{pitch}}[t]=\Delta\theta_{2}^{\mathrm{pitch}}[t]=0$
(Fig. \ref{fig:Geometry}a), the LOS angle between vehicles $\upsilon_{1}$
and $\upsilon_{2}$ is: 
\begin{align}
\overline{\theta}^{\mathrm{LOS}}[t]=\text{tan}^{-1}\bigg(\frac{\bar{h}_{2}-\bar{h}_{1}}{D[t]}\bigg),\label{eq:LOS}
\end{align}
where $D[t]$ indicates the inter-vehicle distance at time $t$. Assuming
that the transmit/receive beams at the two vehicles are steered according
to the nominal LOS direction (\ref{eq:LOS}), if a chassis pitch occurs
(e.g., due to road conditions, electro-mechanical vehicle configurations,
engine vibrations), it causes vertical oscillations of the antenna
arrays which modifies the beam pointing geometry. This effect is highlighted
in Fig. \ref{fig:Geometry}b, where the LOS has changed to 
\begin{align}
\begin{split}\theta^{\mathrm{LOS}}[t] & =\text{tan}^{-1}\bigg(\frac{h_{2}[t]-h_{1}[t]}{D[t]}\bigg)\\
 & \approx\overline{\theta}^{\mathrm{LOS}}[t]-\Delta\theta_{1}^{\mathrm{pitch}}[t]-\Delta\theta_{2}^{\mathrm{pitch}}[t]\label{eq:LOSwithpitch}
\end{split}
\end{align}
due to the rotations $\Delta\theta_{v}^{\mathrm{pitch}}[t]$ at the
two vehicles $v=1,2$. The above change of the LOS elevation angle,
together with the rotation of the two antenna arrays, leads to a change
in the beams' directions which are no more pointing towards the TX-RX
link (black beams in Fig. \ref{fig:Geometry}b). Note that the approximation
in (\ref{eq:LOSwithpitch}) holds since the vibrations (in the order
of few centimeters) are much smaller than the inter vehicle distance
(in the order of several meters) and thus it is $|h_{v}[t]-\bar{h}_{v}|/D[t]\ll1$.

To reduce the misalignment, in the following section we propose an
on-antennas inertial sensor assisted beam tracking solution. Each
array uses its own sensor data and a prediction of the other array
dynamics based on V2V data exchange to track the pitch terms in (\ref{eq:LOSwithpitch})
inside the $\mathrm{BI}$ and dynamically steers its beam according
to the predicted LOS. This technique allows to improve the beam pointing
inside the BI and to avoid too frequent time-consuming BA procedure
by relying only on sensor data and predictive filters, thus increasing
the communication throughput. A visual representation of this concept
is highlighted in Fig. \ref{fig:Geometry}b where the wrongly pointing
black beams have to be steered to point along the LOS link at best
(red beams).

\section{Inertial Sensor Assisted Beam Tracking}

In the inertial sensor aided beam tracking system, a prediction is
made by each transceiver on the behavior of its sensor based on past
dynamics measurements. The prediction is then sent to the transceivers
of the surrounding vehicles, with negligible overhead. Finally, each
transceiver adjusts the beam pointing direction according to its own
sensor data and the information received from the transceivers of the other vehicles. The steps of the proposed technique are detailed in
the following.

\subsection{Strokes prediction}

The basis of beam tracking is a prediction of the stroke dynamics
$h_{v}[t]$ performed at each transceiver $v=1,2$ at the beginning
of the BI. Given the vehicle stroke $h_{v}[t]$, the future process
samples $h_{v}[t+i]$, $i>0$, are predicted locally as a combination of its own past samples. To guarantee
this step, a memory with the recent history of antennas dynamics has to be saved as essential for stroke prediction.

\subsection{V2V communication}

\label{V2V comm}

Once a set of $I$ predicted stroke samples $\hat{h}_{v}[t+i]$, $i=1,\ldots,I,$
has been calculated at transceiver $v$, it is communicated to the
other transceiver over the V2V link and used for beam tracking within
the next BI. Assuming a time division duplex (TDD) protocol for mmWave
communications, where signaling for beam pointing is handled at the
beginning of each frame, we propose a frame structure which reduces
the occurrence of BA procedures and maximizes the payload.

\begin{figure} []
	\centering
	\includegraphics[width=0.7\columnwidth]{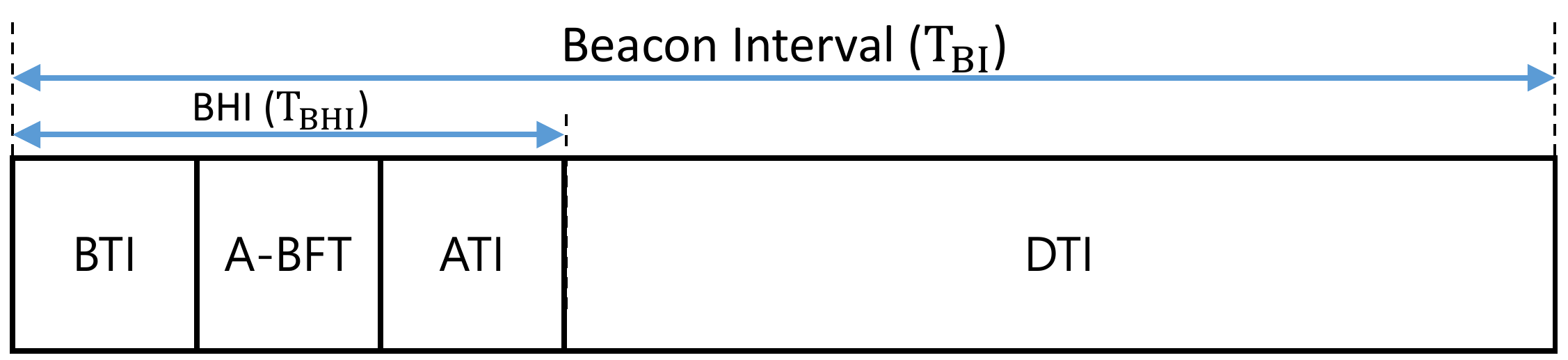}
	\caption{IEEE 802.11ad frame. Besides the slot dedicated to data transfer (DTI), the frame presents an additional payload for signaling information (BHI).} 
	\label{fig:IEEE802.11ad frame}
\end{figure}

We consider as a reference the IEEE 802.11ad/WiGig standard \cite{IEEE802.11ad_Standard,Nitsche2014},
a widely adopted mmWave technology operating in the 60 GHz band mainly
designed for stationary/quasi-stationary applications and here adapted
to the context of V2V communications. As presented in Fig. \ref{fig:IEEE802.11ad frame},
the transmission frame is divided into two main access periods: a
beacon header interval (BHI), dedicated to the exchange of management
information and network announcements, and a data transmission interval
(DTI), where data are transmitted. The BHI consists of three sub-intervals,
the beacon transmission interval (BTI), the association beamforming
training (A-BFT) and the announcement transmission interval (ATI).
Even if the analysis of the frame structure is out of the scope of
this work, the review on BI structure is necessary to detail the BA
procedure. In the IEEE 802.11ad standard, BA consists of two different
phases: a sector level sweep (SLS) and a beam refinement protocol
(BRP) phases. An initial coarse-grained antenna sector configuration
is determined in the SLS and it is further refined during the BRP.
Relying on a contention-based approach, the SLS occurs during BTI
and A-BFT slots, while the BRP can be done in the ATI or DTI. A typical
BI length for stationary, or quasi-stationary, applications is 100
ms. However, for latency-critical tactile-like applications where
short BI lengths are needed, the overhead for time-consuming BA procedures
severely impacts on the communication throughput, as the DTI is significantly
compressed. Considering that in V2X applications, a  $\text{BI}\leq30$
ms is typically considered to face the high mobility \cite{Heath2016b,Mavromatis2017},
a reduction of the BHI signaling is advisable. Indeed, 5G mobile systems (where V2X communication is a primary use case) rely on a 10 ms long radio frame. 

The key idea of the proposed beam tracking approach is to reduce the
signaling for beam alignment so as to shorten the BHI period and extend
the DTI, with benefits in transmission efficiency. By exploiting inertial
sensor assisted predictive models, the best beam pair search between
transmitter and receiver is substituted by beam pointing based on
predicted stroke information $\hat{h}_{v}[t+i]$, with negligible
additional signaling payload (i.e., the transmission of the predicted
samples only). Denoting with $\mathrm{\text{T}_{S}}$ the overall BI overhead
due to signaling information and with $\mathrm{\text{T}_{BA}}$ the overhead
portion for the BA procedure (i.e., SLS and BRP), we can define the
transmission efficiency as the ratio of the time interval dedicated
to data transfer over the BI duration $\text{T}_\text{BI}$, as follows: 
\begin{align}
\eta=\left\{ \begin{array}{ll}
1-\frac{\mathrm{\text{T}_{S}}+\mathrm{\text{T}_{BA}}}{\text{T}_\text{BI}}\approx1-\frac{\text{T}_{\text{BHI}}}{\text{T}_\text{BI}} & \text{conventional BA protocol,}\vspace{0.15cm}\\
1-\frac{\mathrm{\text{T}_{S}}}{\text{T}_\text{BI}} & \text{sensor aided beam tracking.}
\end{array}\right.\label{eq: frame ratio}
\end{align}
Considering that BA is the dominant signaling payload, the $\mathrm{\text{T}_{BA}}$
removal by the proposed technique leads to significant benefits in
terms of transmission efficiency, as discussed in Sec. \ref{Results}
by numerical analysis.

\subsection{Beam tracking}

As BA is essential for mmWave directional communications, the choice
of the optimal pointing direction is crucial to guarantee reliable
communications. This issue is far more critical when dealing with
mMIMO systems, where the beamwidth is narrow. Focusing for convenience
on a single BI period, at time instant $t=0,1,\ldots,\mathrm{T_{BI}}$,
vehicle ${v}_{1}$ should direct the beam towards the ideal pointing
angle $\theta_{1}^{\mathrm{point}}[t]$ given by (see Fig. \ref{fig:Geometry}b):
\begin{align}
\theta_{1}^{\mathrm{point}}[t]=-\Delta\theta_{1}^{\mathrm{pitch}}[t]+{\theta}^{\mathrm{LOS}}[t],\label{eq:pointing angle IDEAL}
\end{align}
where $\Delta\theta_{1}^{\mathrm{pitch}}[t]$ accounts for the rotation
of the antenna array at vehicle ${v}_{1}$. However, due to lack on
real-time knowledge of the dynamics of the communication device at vehicle $v_{2}$, the pointing
angle of $v_{1}$ is 
\begin{align}
\hat{\theta}_{1}^{\mathrm{point}}[t]=-\Delta\theta_{1}^{\mathrm{pitch}}[t]+\hat{\theta}^{\mathrm{LOS}}[t],\label{eq:pointing angle}
\end{align}
where $\hat{\theta}^{\mathrm{LOS}}[t]$ is the estimate of the elevation
angle of the LOS path connecting the two V2V devices (see Fig. 3):
\begin{align}
\hat{\theta}^{\mathrm{LOS}}[t]=\left\{ \begin{array}{ll}
\text{tan}^{-1}\bigg(\frac{h_{2}[t_{\text{BA}}]-h_{1}[t]}{D[t]}\bigg)+n_{\text{BA}} & \text{conventional BA protocol,}\vspace{0.15cm}\\
\text{tan}^{-1}\bigg(\frac{\hat{h}_{2}[t]-h_{1}[t]}{\hat{D}[t]}\bigg) & \text{sensor aided beam tracking.}
\end{array}\right.\label{eq:3LOS}
\end{align}
Here, $\hat{D}[t]=D[t]+n_{d}[t]$ denotes the measured inter-vehicle distance
modeled with $n_{d}[t]\sim\mathcal{N}(0,\sigma_r^{2})$, where $\sigma_r$ represents the measurement accuracy provided, for example, by a radar system.
 The additional
term $n_{\text{BA}}$ takes into account for misalignments occurring
while searching the best beam pair in conventional BA protocol. It
is modeled as a random variable uniformly distributed as $n_{\text{BA}}\sim\mathcal{U}(-\theta_{3\text{dB}},+\theta_{3\text{dB}})$,
where $\theta_{3\text{dB}}$ denotes the antenna resolution. For a
uniform linear array (ULA) the resolution is evaluated as the $3$-dB
beamwidth in angle-space, which is typically related to the number
of antenna elements $N$ by the approximation $\theta_{3\text{dB}}\approx\sfrac{0.866}{N}$
rad \cite{OrfanidisBoook}. From \eqref{eq:3LOS}, it is evident that
a perfect awareness of the self dynamics is not sufficient to get
a perfect pointing since $\hat{\theta}^{\mathrm{LOS}}[t]$ depends
also on ${v}_{2}$ device dynamics.

\begin{figure}[]
	\centering
	\includegraphics[width=0.7\columnwidth]{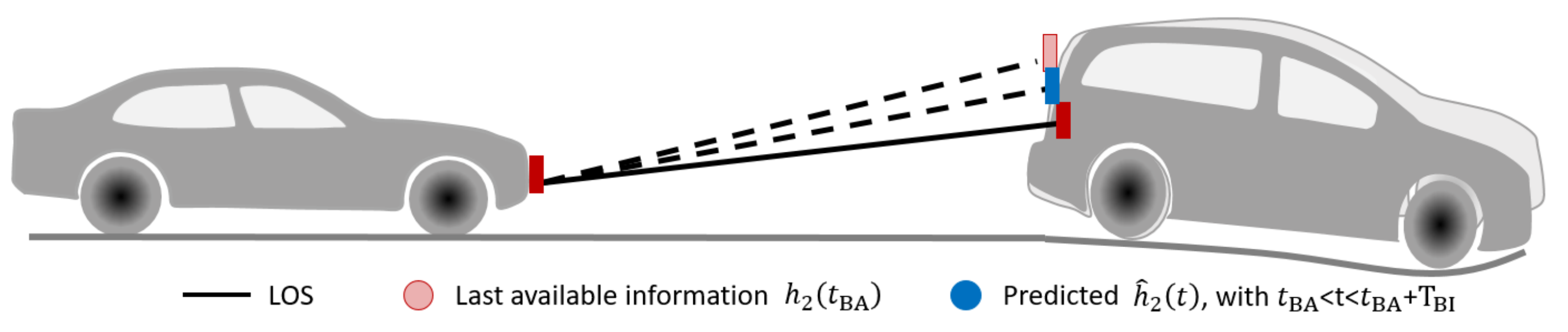}
	\caption{Beam tracking effect: a prediction on the stroke dynamics based on an inertial sensor in the communication device allows to properly point the beam, reducing misalignments.}
	\label{fig:beam tracking}
\end{figure}

In traditional BA systems transceivers are unaware of the other transceiver
dynamics inside the BI and vehicles can only use the information available
at the time BA is performed (i.e., at time $t=t_{\text{BA}}$). Using
side information on predicted dynamics exchanged at the beginning
of the BI as in the proposed approach the term $\hat{\theta}^{\mathrm{LOS}}[t]$
can be updated all over the BI, i.e. $\forall t\in(t_{\text{BA}},t_{\text{BA}}+\text{T}_{\text{BI}})$.
In this way, the transceiver ${v}_{1}$ can optimize its pointing
angle $\hat{\theta}_{1}^{\mathrm{point}}[t]$ over time, approaching
the ideal condition of perfect LOS communication, as shown in Fig.
\ref{fig:beam tracking}. The benefit of the proposed system is twofold,
i.e. an increased transmission efficiency $\eta$ due to the signaling
reduction (appreciated especially for short BI) and a beam tracking
gain provided by dynamics prediction (for longer BI).

\section{V2V mmWave Channel Modeling}

In the following, we evaluate the impact of the mismatch between the
true and estimated beam-pointing, ${\theta}_{v}^{\mathrm{point}}[t]$
and $\hat{\theta}_{v}^{\mathrm{point}}[t]$, at vehicles $v=1,2$
on the V2V transmission capacity. We assume a mmWave mMIMO LOS and narrow band communication
link between the two vehicles, with a ULA of $N$ antennas at both
vehicles.

The received power at time $t$ is modeled as: 
\begin{align}
P_{\text{rx}}^{\text{dB}}[t]=P_{\text{tx}}^{\text{dB}}+G_{1}[t]+G_{2}[t]-PL^{\text{dB}}[t]\label{eq:Prx}
\end{align}
where $P_{\text{tx}}^{dB}$ is the transmitted power, $G_{1}[t]$
and $G_{2}[t]$ are the ULA antenna gains at the two vehicles and
$PL^{\text{dB}}[t]$ the channel path loss. The latter is defined
as: 
\begin{align}
PL^{\text{dB}}[t]=20\text{log}_{10}\bigg(\frac{4\pi}{\lambda}\bigg)+10\kappa\text{log}_{10}D[t]+\chi_{sh},\label{eq:path loss}
\end{align}
where $\lambda$ is the carrier wavelength, $\kappa$ the path loss
exponent and $\chi_{sh}$ is the log-normal distributed shadowing,
$\chi_{sh}\sim\mathcal{N}(0,\sigma_{\text{dB}}^{2})$ \cite{Akdeniz2014}.

The main vehicle dynamics-dependent parameters in \eqref{eq:Prx}
are the antenna array gains $G_{v}[t]$, $v=1,2$, which depend on
the beam-pointing mismatch that for ideal uniform linear array is:

\begin{align}
\begin{split}G_{v}[t]= & G\bigg(\hat{\theta}_{v}^{\mathrm{point}}[t]\bigg|\theta_{v}^{\mathrm{point}}[t]\bigg)\\
= & \frac{1}{N}\bigg|\mathbf{a}\big(\hat{\theta}_{v}^{\mathrm{point}}[t]\big)^{\text{H}}\mathbf{a}\big(\theta_{v}^{\mathrm{point}}[t]\big)\bigg|^{2}\\
= & \bigg|\frac{\text{sin}\big[\pi\big(\text{sin}(\hat{\theta}_{v}^{\mathrm{point}}[t])-\text{sin}(\theta_{v}^{\mathrm{point}}[t])\big)\sfrac{N}{2}\big]}{N\text{sin}\big[\pi\big(\text{sin}(\hat{\theta}_{v}^{\mathrm{point}}[t])-\text{sin}(\theta_{v}^{\mathrm{point}}[t])\big)/2\big]}\bigg|^{2}.\label{eq:antennagain}
\end{split}
\end{align}
$\mathbf{a(\theta)}=[a_{1}(\theta),...,a_{N}(\theta)]^{\mathrm{T}}\in\mathbb{R}^{N\times1}$
is the steering vector modeling the antenna array response to a direction
$\theta$ that for a half-wavelength spacing between the antenna elements
is: 
\begin{align}
a_{n}(\theta)=h_{n}\text{e}^{-j\pi(n-1)\text{sin}\theta}, 1\leq n\leq N.
\end{align}
The term $h_{n}$ is introduced to account for the mismatches of the
transceivers radio frequency (RF) circuits at the antenna elements
and array calibration errors due to hardware themselves and properties
of surrounding environments (e.g., temperature and moisture). The
RF mismatch is modeled as 
$h_{n}=\rho_{n}\text{e}^{j\psi_{n}},$ 
with log-normally distributed amplitude $10\mathrm{log}(\rho_{n})\sim\mathcal{N}\big(0,\delta^{2}\big)$
and uniformly distributed phase $\psi_{n}\sim\mathcal{U}\big(-\phi,+\phi\big)$
\cite{Huawei2009,Björnson2014}.

The noise power at the receiver is evaluated as 
\begin{align}
P_{\text{noise}}^{\text{dB}}=N_{fl}+10\text{log}_{10}B+NF,\label{eq:Pnoise}
\end{align}
where $N_{fl}$ is the noise floor ($N_{fl}=-174$ $\text{dBm/Hz}$),
$B$ is the system bandwidth and $NF$ is the noise figure. The signal-to-noise
ratio (SNR) is finally evaluated as the ratio between the receive
and noise power as: 
\begin{align}
\text{SNR}[t]=\frac{P_{\text{rx}}[t]}{P_{\text{noise}}[t]},\label{eq:SNR}
\end{align}

Recalling the definition of $\eta$ in Sec. \ref{V2V comm}, the maximum
achievable data rate in the V2V link between ${v}_{1}$ and ${v}_{2}$
in a single BI is obtained as: 
\begin{align}
R_{12}[t]=\eta B\text{log}\big(1+\text{SNR}[t]\big).\label{eq:data rate}
\end{align}
The above V2V data rate is used in the following section to assess
the performance of the proposed inertial sensor aided beam tracking
technique.

\section{Numerical Results}

\label{Results} 
\begin{figure}
	\centering
	\includegraphics[width=0.7\columnwidth]{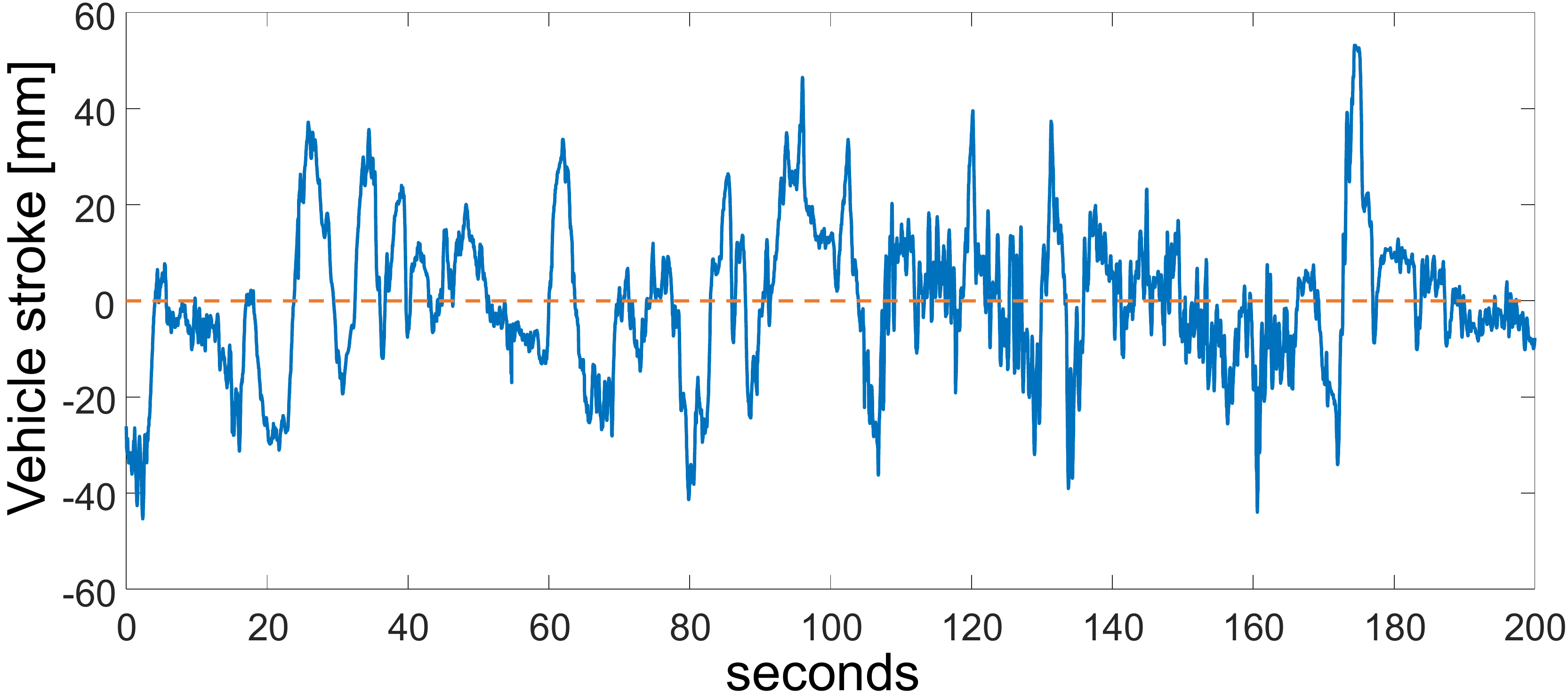}
\caption{Real measurements of vehicle strokes provided by the inertial sensor.}
	\label{fig:real stroke} 
\end{figure}
\begin{figure}
	\centering
	\includegraphics[width=0.7\columnwidth]{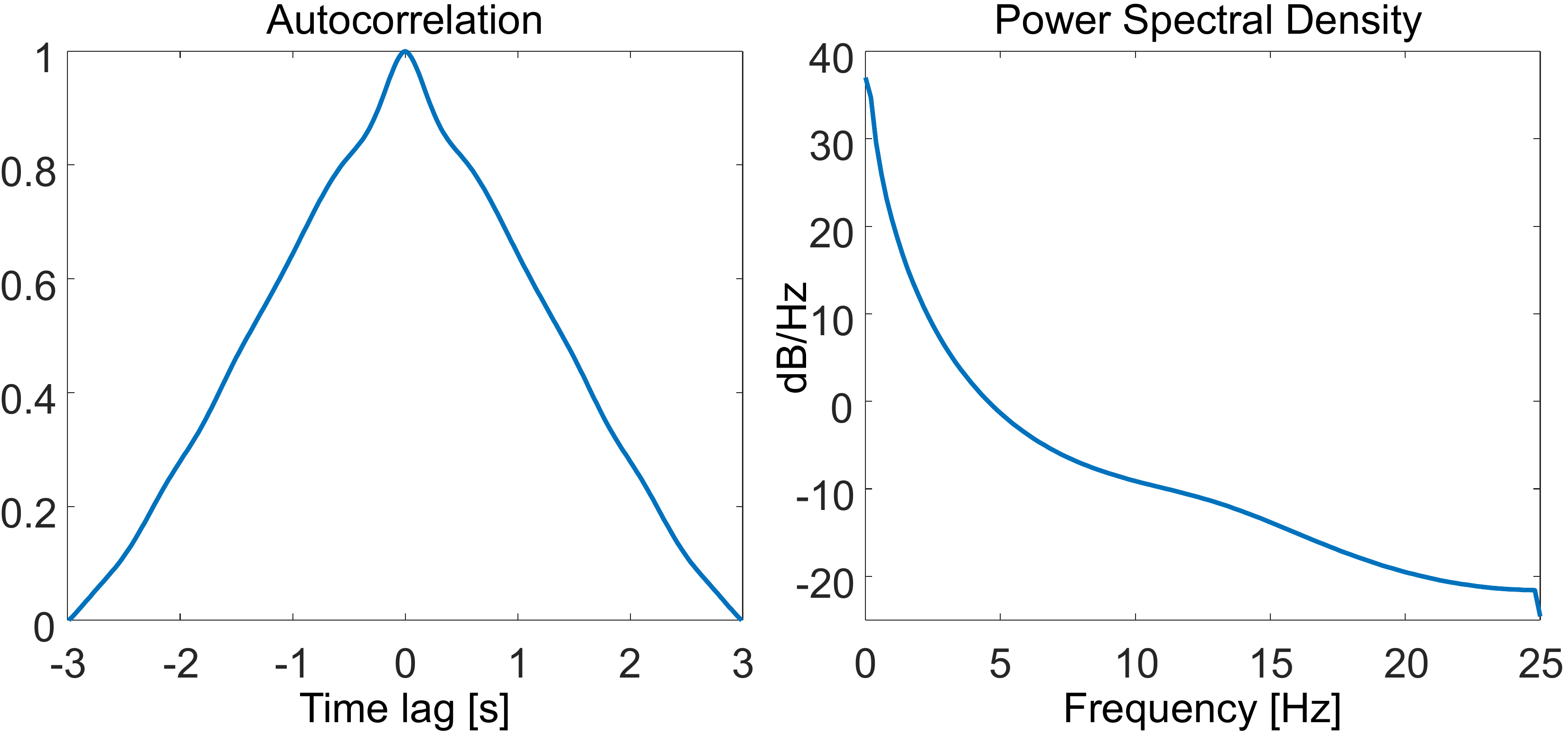}
	\caption{Vehicle stroke properties: autocorrelation and power spectral density.}
	\label{fig: autocorrelation and psd}
\end{figure}

In this section we assess the performance of the proposed inertial
sensor aided beam tracking technique based on real measurements of
vehicle strokes. A measurement campaign has been carried out for an
acquisition interval of $200$ s, where data have been gathered with
a sampling frequency of $50$ Hz. The data have been collected on
a high-end passenger car (sedan), using the accelerometers transformed
to vertical component only. These data are plotted in Fig. \ref{fig:real stroke},
while their statistical properties, namely the normalized autocorrelation
and the power spectral density, are presented in Fig. \ref{fig: autocorrelation and psd}.
The vehicles' strokes $h_{1}[t]$ and $h_{2}[t]$ are modeled as autoregressive
AR(10) processes with parameters calibrated on the real measurements.
To match the strokes resolution to the BI duration, an interpolation
factor of 10 is applied to $h_{1}[t]$ and $h_{2}[t]$ to have a resolution
in time of $\Delta t=2\text{ms}$. The following results
are obtained assuming a constant V2V distance of $D(t)=5\text{m}$,
which is a typical value in platooning applications. Its estimate
$\hat{D}(t)$ is updated by the vehicles every $200$ $\text{ms}$.
The simulation parameters are summarized in Tab. \ref{tab:Simulation Parameters}.

\begin{table}
	\centering
	\caption{Simulation Parameters}
	\label{tab:Simulation Parameters}
	\begin{footnotesize}
		\begin{tabular}{l c | l c }
			Parameter    & Value & Parameter & Value \\
			\hline  
			Length $\ell_1$ & 4.5 \text{m}   & 
			Length $\ell_2$ & 5 \text{m}    \\
			Height $\bar{h}_1$ & 0.5 \text{m}   & 
			Height $\bar{h}_2$ & 1 \text{m}    \\
			Overhead $\mathrm{\text{T}_{BA}}$ & $1.9$ \text{ms} &
			Overhead $\mathrm{\text{T}_S}$   & $ 0.1 $ \text{ms} \\
			Predictor length $p$ & 11  & V2V distance $D$ & 5 \text{m} \\Carrier Frequency & 60 \text{GHz}     &
			System Bandwidth $B$ & 2.16 \text{GHz} \\
			Tx Power $P_{tx}$ & 1 \text{dBm}&
			Path Loss Exponent $\kappa$ & 2 \\
			Shadowing $\sigma_{\text{dB}}$ &  $5.8$ $ \text{dB}$   &
			Noise Factor $NF$ & 6 \text{dB}	 \\
			RF mismatch $\delta_{\text{dB}}$  & 1 \text{dB} &
			RF phase mismatch $\phi$  & 3\textdegree      \\
			\hline        
		\end{tabular}
	\end{footnotesize}
\end{table}

\begin{figure*}[]
	\centering
	\includegraphics[width=1\linewidth, height=\textheight, keepaspectratio]{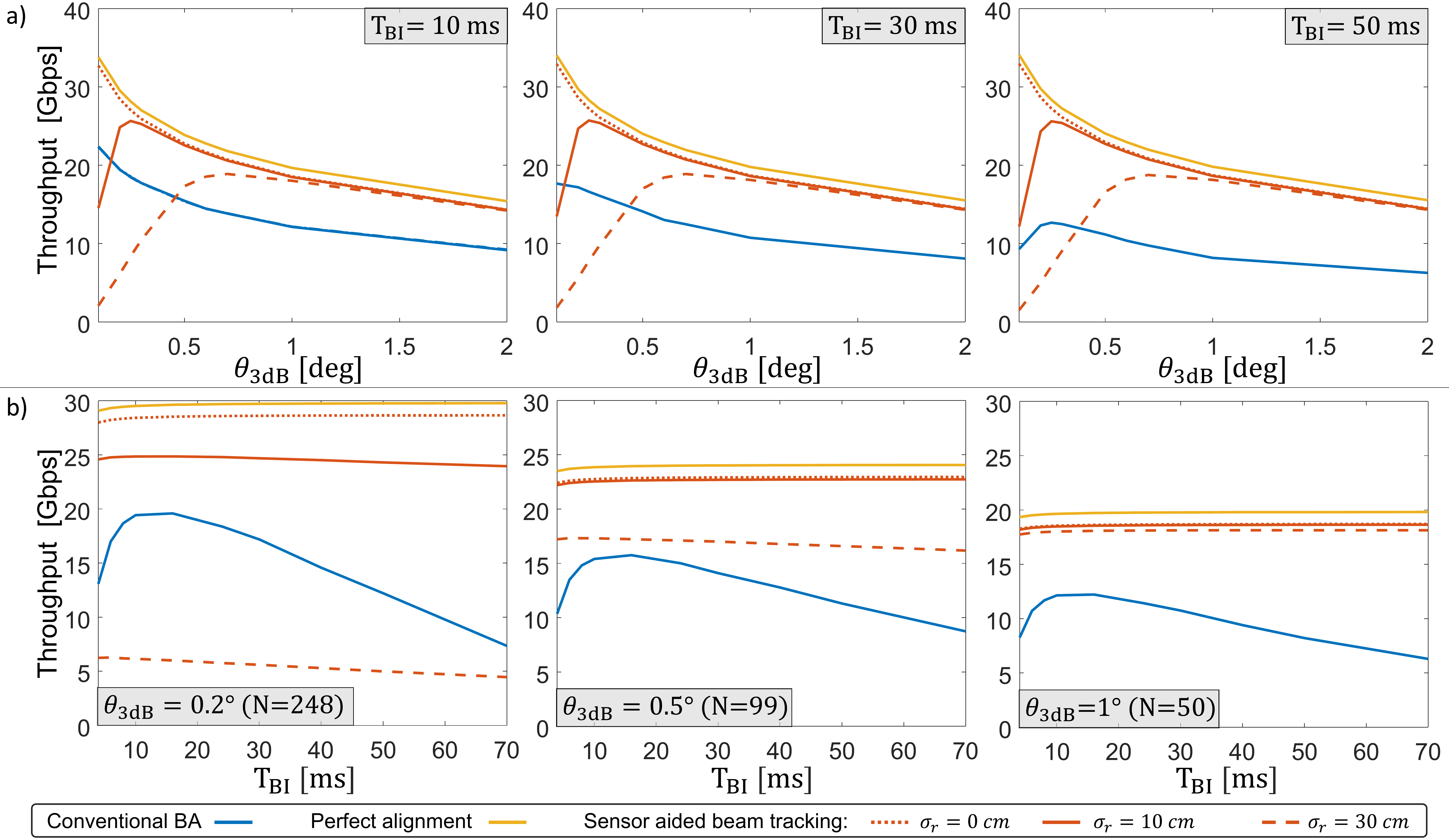}
	\caption{V2V communication throughput comparison. The inertial sensor aided beam tracking is compared to conventional BA protocol and to an ideal communication with perfect alignment. a) The V2V throughput is plotted versus the BI duration $\text{T}_{\text{BI}}$ for three different beamwidth. b) The V2V throughput is presented versus the beamwidth for three different $\text{T}_{\text{BI}}$.}
	\label{fig:results_rate_vs_BI}
\end{figure*}

We compare the performance of the proposed inertial sensor aided beam
tracking approach with respect to conventional BA techniques, taking
the IEEE 802.11ad standard as a reference protocol. We also consider
as upper bound the performance of an ideal V2V communication system
with perfect BA based on exact knowledge of the geometrical parameters
at each time $t$). The maximum achievable V2V data rate (in Gbps)
is used as a performance indicator and it is presented versus the
array resolution $\theta_{3\text{dB}}$ in Fig. \ref{fig:results_rate_vs_BI}-(a)
and versus the BI duration $\text{T}_{\text{BI}}$ in Fig. \ref{fig:results_rate_vs_BI}-(b).
The results are presented for three different values of radar accuracy
$\sigma_r$, which impacts the performance only in the sensor aided
technique. In the throughput evaluation, we consider a frame to be
erroneously received if $\exists t\in$ BI s.t. $\text{SNR}[t]<\text{SNR}_{\text{ideal}}[t]-6\text{dB}$.

Results in Fig. \ref{fig:results_rate_vs_BI} indicate that sharp
beams (or equivalently small $\theta_{3\text{dB}}$) increase the
throughput as the power is concentrated in a narrower angular space,
however they are more sensitive to the variation of vehicle dynamics
and to the inter-distance accuracy in the V2V distance estimate. A
sensor aided beam tracking method can closely approach the performance
of an ideal communication system with perfect alignment if the V2V
distance is perfectly estimated (i.e., for $\sigma_r=0$ $\text{cm}$).
On the other hand, degrading effects due to poor ranging systems (e.g.,
for $\sigma_r=30$ $\text{cm}$) occur at small $\theta_{3\text{dB}}$.
Considering that a typical accuracy in V2V application is $\sigma_{r}=10$
$\text{cm}$, we can conclude that a sensor aided beam tracking system
can provide a higher V2V throughput for $\theta_{3\text{dB}}>0.2$\textdegree $ $
for any $\text{T}_{\text{BI}}$ with respect to conventional BA protocols. Note that for long BI ($\text{T}_{\text{BI}}=50$ $\text{ms}$), the use of sharp beams reduces the throughput even for conventional BA systems, as vehicles are likely to be in misalignment conditions.

\section{Conclusions}

In this paper, we developed an innovative solution for beam tracking
in V2V communications based on side information of antenna array
dynamics. The proposed inertial sensor aided beam tracking system
allows to continuously steer the beam while transmitting, it reduces
the signaling payload of the communication and, lastly, it avoids
a time-consuming search of the best beam pair between transmitter
and receiver. Results demonstrated an increase in the V2V data rate
with respect to conventional BA protocols for any frame duration and
for array resolution $\theta_{3\text{dB}}>0.2$\textdegree $ $  when typical
values of ranging accuracy (e.g., $\sigma_{r}=10$ $\text{cm}$) are
used. The proposed methodology can be extended to handle the communication between a vehicle and a fixed station, as for V2I. 

 \bibliographystyle{IEEEbib}
\bibliography{Bibliography}

\end{document}